\documentclass[useAMS,usenatbib]{mnras}

\usepackage{graphics}
\usepackage{amssymb}
\usepackage{amsfonts}
\usepackage{wasysym}
\usepackage{epsfig}

\usepackage{longtable}
\usepackage{times}
\usepackage{mathptmx}
\usepackage[usenames,dvipsnames]{color}

\newcommand{\s}{\scriptsize}

\newcommand{\XMM}{{\em XMM-Newton}}

\title[The magnetosphere of HR\,7355] 
{The detection of variable radio emission from the fast rotating magnetic 
hot B-star HR\,7355 and evidence for its X-ray aurorae}
\author[P. Leto et al.]
{P. Leto$^{1}$ \thanks{E-mail: pleto@oact.inaf.it},
C. Trigilio$^{1}$,
L. Oskinova$^{2}$,
R. Ignace$^{3}$,
C. S. Buemi$^{1}$,
G. Umana$^{1}$,
\newauthor A. Ingallinera$^{1}$,
H. Todt$^{2}$,
F. Leone$^{4}$
\\
$^{1}$INAF - Osservatorio Astrofisico di Catania, Via S. Sofia 78, 95123 Catania, Italy\\
$^2$Institute for Physics and Astronomy, University Potsdam, 14476 Potsdam, Germany\\
$^3$Department of Physics \& Astronomy, East Tennessee State University, Johnson City, TN 37614, USA\\
$^4$Universit\'{a} degli studi di Catania, Via S.Sofia 78, I-95123 Catania, Italy
}
\begin{document}

\date{}

\pagerange{\pageref{firstpage}--\pageref{lastpage}} \pubyear{}

\maketitle

\label{firstpage}

\begin{abstract}
In this paper we investigate the multiwavelengths properties 
of the magnetic early B-type star HR\,7355. We present its radio 
light curves at several frequencies, taken with the Jansky Very
Large Array, and X-ray
spectra, taken with the \XMM\ X-ray telescope.  
Modeling of the radio light curves for the Stokes I and V
provides a quantitative analysis of the HR\,7355 magnetosphere.
A comparison between HR\,7355 and
a similar analysis for the Ap star CU\,Vir,
allows us to study how the different physical parameters of the two stars affect
the structure of the respective magnetospheres where the non-thermal electrons originate.
Our model includes a cold thermal plasma component that accumulates
at high magnetic latitudes that influences the radio regime, but does
not give rise to X-ray emission.
Instead, the thermal X-ray emission
arises from shocks generated by wind stream collisions close 
to the magnetic equatorial plane.
The analysis of the X-ray spectrum of HR\,7355 also suggests
the presence of a non-thermal radiation.
Comparison between the spectral index of the power-law X-ray energy distribution
with the non-thermal electron energy distribution indicates
that the non-thermal X-ray component could be the auroral signature of
the non-thermal electrons that impact the stellar surface, the same non-thermal
electrons that are responsible for the observed radio emission.
On the basis of our analysis, we suggest a novel model that simultaneously 
explains the X-ray and the radio features of HR\,7355 and is likely 
relevant for magnetospheres of other magnetic early type stars.
\end{abstract}


\section{Introduction}

Stellar magnetism at the top of the main sequence is not typical, but
neither is it an extremely rare phenomenon. 
In fact about 10\% of the OB-type stars display strong and stable magnetic fields
\citep{grunhut_etal12b,fossati_etal15}.
The hot magnetic stars are mainly characterized as oblique magnetic rotators (OMR):
a dipolar magnetic field
topology with field axis misaligned with respect to the rotation
axis \citep{babcock49,stibbs50}.  The
existence of such well-ordered magnetic fields are a cause of inhomogeneous
photospheres, giving rise to observable photometric, spectroscopic
and magnetic variability that can be explained  in the framework of
the OMR. Early type magnetic stars  are sufficiently 
hot to produce a radiatively driven stellar wind, that in the presence of their
large-scale magnetic fields may be strongly aspherical.  The wind plasma 
accumulates at low magnetic latitudes (inner magnetosphere), 
whereas it can freely propagate along directions near the magnetic poles \citep{shore87,leone93}.
Observable signatures of plasma trapped inside 
stellar magnetospheres can be recognized in the UV spectra
\citep{shore_etal87,shore_brown90} and in the H$\alpha$ line
\citep{walborn74}.
The interaction of a radiatively driven wind with the stellar
magnetosphere has been well studied. As examples, see 
\citet*{poe_etal89} for an application of the \citet{weberdavis}
model to radiatively driven winds,
wind compression models with magnetic fields (WCFields) \citep*{ignace_etal98},
and the magnetically torqued disk (MTD) model \citep{cassinelli_etal02}. 
Such interaction 
causes an accumulation of hot material close to the magnetic
equatorial plane, as described by the magnetically confined wind
shock (MCWS) model \citep{babel_montmerle97}.
The wind plasma arising from the two opposite hemispheres collides
close to the magnetic equatorial plane, shocking the plasma
to radiate X-rays
(\citealp{ud-doula_naze15} has made a recent review of the X-ray
emission from magnetic hot stars).
In the presence of a strong magnetic field, the 
stellar wind plasma is confined within the stellar magnetosphere and
forced to rigidly co-rotate with the star
\citep{townsend_owocky05,ud-doula_etal06,ud-doula_etal08}.
The prototype of a rigidly rotating magnetosphere (RRM)
is $\sigma$\,Ori\,E \citep{groote_hunger97,townsend_etal05}.
Evidence of circumstellar matter bound to the strong stellar magnetic field 
has been reported in a few other cases
\citep{leone_etal10,  
bohlender_monin11,  
grunhut_etal12a, 
rivinius_etal13, 
hubrig_etal15, 
sikora_etal15}. 

The stellar rotation plays an important role in establishing the size of
the RRM, and the density
of the plasma trapped inside.  
The rotation works in opposition to the gravitational infall
of the magnetospheric plasma, leading to a large centrifugal
magnetosphere (CM) \citep{maheswaran_cassinelli09,petit_etal13}.

The existence of a CM filled by stellar wind material
is a suitable condition to give rise to non-thermal radio continuum
emission that was first measured for 
peculiar magnetic B and A stars  \citep*{drake_etal87,linsky_etal92,leone_etal94}.
In accord with the OMR model, their radio
emission is  cyclic owing to stellar rotation
\citep{leone91,leone_umana93}, suggestive of optically thick
emission arising from a stable RRM. 
The radio emission features 
are characterized by a simple dipolar magnetic field topology
and have been successfully reproduced using a 3D model that computes
the gyrosynchrotron emission 
\citep{trigilio_etal04,leto_etal06}.

In this model, the non-thermal electrons
responsible for the radio emission originate in magnetospheric regions
far from the stellar surface, where the kinetic energy density of
the gas is high enough to brake the magnetic field lines forming
current sheets. These regions are the sites where the mildly
relativistic electrons originate.  In the magnetic equator, at
around the Alfv\'en radius, there is a transitional magnetospheric
``layer'' between the inner confined plasma and the escaping wind.
Energetic electrons that recirculate through this layer back to the
inner magnetosphere radiate radio by the gyrosynchrotron emission
mechanism.  

The analysis of radio emissions from magnetic early-type stars
is a powerful diagnostic tool for the study of the topology
of their magnetospheres.  The radio radiation at different frequencies
probes the physical conditions of the stellar magnetosphere
at different depths, even  topologies are complex \citep{leto_etal12}.
Hence, the radio emission  of the hot magnetic stars
provides a favored window to study 
the global magnetic field topology, the spatial stratification of the thermal electron density, 
the non-thermal electron number density, and interactions between stellar 
rotation, wind, and magnetic field.
In fact, the above physical parameters can be derived by comparing
the multi-wavelength radio light curves, for the total and circularly polarized flux density, 
with synthetic light curves using our 3D theoretical model. 
It is then possible, to study how  such stellar properties as rotation, wind, magnetic 
field geometry affect the efficiency of the electron 
acceleration mechanism.

In particular,
it is important to apply the radio diagnostic techniques on a sample of 
magnetic early-type stars that differ
in their stellar rotation periods, magnetic field strengths, and field geometries.
To this end, we conducted a radio survey of a representative sample of hot magnetic stars 
using the Karl G. Jansky Very Large Array (VLA).
These stars probe different combinations of source parameters owing to their 
different physical properties.
This paper presents the first results of this extensive study.

Here we present the analysis of the radio emission from the fast rotating, hot 
magnetic star HR\,7355.  We were able to
reproduce multi-wavelength radio light curves for the total and the
circularly polarized flux density.  The model simulation of the
radio light curves, along with a simulation of the X-ray spectrum
of HR\,7355, are used to significantly constrain the physical
parameters of its stellar magnetosphere. On this basis, we
suggest a scenario that simultaneously explains the
behavior of HR\,7355 at both radio and X-ray wavelengths.

In Section\,\ref{sec:uv} we briefly introduce the object of this study, 
HR\,7355.  The observations used in our analyses are 
presented in Section\,\ref{sec:obs}. The radio properties of HR\,7355 are 
discussed in detail in Section\,\ref{sec:radio}. Section\,\ref{model} 
describes the model, while stellar magnetosphere is presented in 
Section\,\ref{sec:magnet}. Analysis of X-ray emission of HR\,7355 is 
provided in Section\,\ref{sec_xray}. The considerations on auroral 
radio emission in HR\,7355 are given in section Section\,\ref{sec:aur}, while 
Section\,\ref{sec:sum} summarizes the results of our work.

\section{Magnetic early B-type star HR\,7355}
\label{sec:uv}

The early-type main sequence star (B2V) HR\,7355 (HD\,182180) is
characterized by a surface overabundance of helium
\citep{rivinius_etal08}.  This star evidences also a very strong
and variable magnetic field \citep{oksala_etal10,rivinius_etal10}.
The magnetic curve of HR\,7355  changes polarity twice per period,
and was modeled in the framework of the OMR by a mainly
dipolar field, with the magnetic axis significantly misaligned with respect
to the rotation axis.

Among the class of the magnetic early-type stars, HR\,7355 is an
extraordinarily fast rotator. Only the B2.5V type star HR\,5907
\citep{grunhut_etal12a} has a shorter rotation period.  The rotation
period of HR\,7355 ($\approx 0.52$ days) sets it close to the
point break-up, giving rise to a strong deformation from spherical
\citep{rivinius_etal13}.  The main stellar parameters of HR\,7355
are listed in Table~\ref{par_star}.

HR\,7355 hosts a strong and steady magnetic field,
indicating the existence of a co-rotating magnetosphere \citep{rivinius_etal13}
suitable for giving rise to non-thermal radio continuum emission.
HR\,7355 has a flux density at 1.4 GHz of $7.9 \pm 0.6$ [mJy] as listed by 
the NVSS \citep{condon_etal98}.
At the tabulated stellar distance, we estimate a radio luminosity of 
$\approx 5\times 10^{17}$ [erg s$^{-1}$ Hz$^{-1}$], making HR\,7355 
one of the most luminous magnetic hot stars at radio wavelengths.
Thus, HR\,7355 is an ideal target to study the effects 
of fast rotation and the high magnetic field strength for magnetospheric 
radio emissions.

\begin{table}
\caption[ ]{Summary of stellar parameters for HR\,7355}
\label{par_star}
\footnotesize
\begin{tabular}[]{lccr}
\hline
\hline
Parameter                                                    &Symbol                               &                                                & ref.      \\
\hline
Distance [pc]                                                 &$D$                                    & 236                                       & {\scriptsize 1}  \\
Reddening [mag]                                         &$E(B-V)$                           & 0.065                                       & {\scriptsize 1}  \\
Mass [M$_{\odot}$]                                      &$M_{\ast}$                        & 6                                          & {\scriptsize 1}  \\
Equatorial Radius [R$_{\odot}$]                      &$R_{\ast}$                         & 3.69                                & {\scriptsize 1}  \\
Photospheric Temperature [K]                  &$T_\mathrm{phot}$          & 17500                                  & {\scriptsize 1}   \\
Rotational Period [days]                            &$P_\mathrm{rot}$              & 0.5214404                            & {\scriptsize 2}\\
Polar Magnetic Field [Gauss]                   &$B_\mathrm{p}$                 & 11600                                 &  {\scriptsize 1} \\
Rotation Axis Inclination [degree]           &$i$                                        & 60                                         & {\scriptsize 1} \\
Magnetic Axis Obliquity [degree]            &$\beta$                                 &    75                                        & {\scriptsize 1}     \\
\hline
\end{tabular}
\begin{list}{}{}
\item[References:]
 {\scriptsize (1)} \citealp{rivinius_etal13};
 {\scriptsize (2)} \citealp{oksala_etal10};
 \end{list}
\end{table}

To obtain information on the stellar wind parameters of HR\,7355 
we retrieved its archival UV spectra (sp39549, sp39596) obtained by 
the International Ultraviolet Explorer (IUE). These UV spectra were
analyzed by means of non-LTE iron-blanketed model atmosphere PoWR, 
which treats the photosphere as well as the the wind, and also accounts 
for X-rays \citep{graf2002, hg2003, Sander2015}. Already the first 
inspection of the spectra reveals a lack of asymmetric line profiles 
which would be expected for spectral lines formed in a stellar wind. 
In fact, \citet{rivinius_etal13} were able to fit the IUE spectra of 
HR\,7355 with a static model that does not include stellar wind, showing 
that its contribution must be small, and at best only upper limit could be 
obtained by UV spectral line modeling.

\begin{figure}
\resizebox{\hsize}{!}{\includegraphics{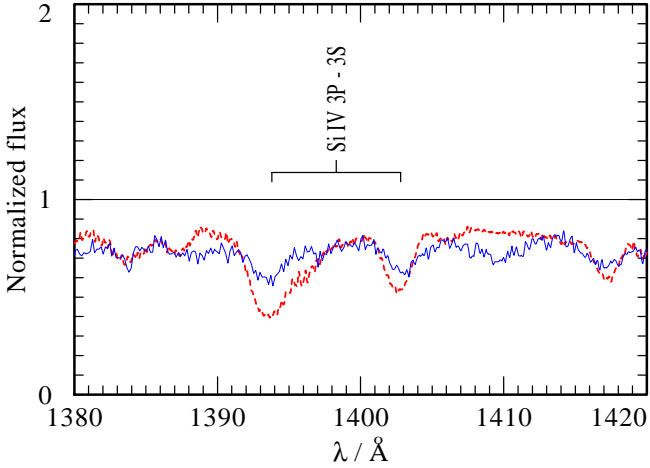}}
\caption{Detail of the IUE observation (blue solid line) vs. a PoWR model
(red solid line) with $T_{\rm eff}=15.7\,kK$, 
$\log (\dot{M}) = -11$ 
and $v_\infty = 500\,$km\,s$^{-1}$. 
The synthetic spectrum was calculated for a rotating atmosphere with 
$v \sin i = 320\,$km\,s$^{-1}$ as described in \citet{Shenar2014}.
}
\label{fig:siiv}
\end{figure}

We attempted to estimate the upper limit for the mass-loss rate that would be 
still consistent with the IUE observations. We found that in our models for the 
stellar parameters given by \citet{rivinius_etal13} and the assumed mass-loss 
rate of about $\dot{M}=10^{-10}\,M_\odot$\,yr$^{-1}$ only the Si\,{\sc iv} 
resonance line is sensitive to the mass-loss rate (see Fig.\,\ref{fig:siiv}). 
As spherical symmetry is assumed for the PoWR models while 
\citet{rivinius_etal13} found different temperatures for the pole and equator 
regions, we can give only a rough estimate for the mass-loss rate. For the lower 
temperature of $15.7\,$kK at the pole region the Si\,{\sc iv} resonance line is 
weaker than for the higher temperatures of the equator region. Therefore we 
infer the upper limit of the mass-loss rate for the lower temperature and find a 
value of about
$\dot{M} < 10^{-11}\,M_\odot$\,yr$^{-1}$ for a spherical symmetric smooth wind 
to be consistent with the IUE observation. We also checked for the effect of
X-rays via super-ionization but did not find a major impact on the Si\,{\sc iv}
resonance line.

\begin{figure*}
\resizebox{\hsize}{!}{\includegraphics{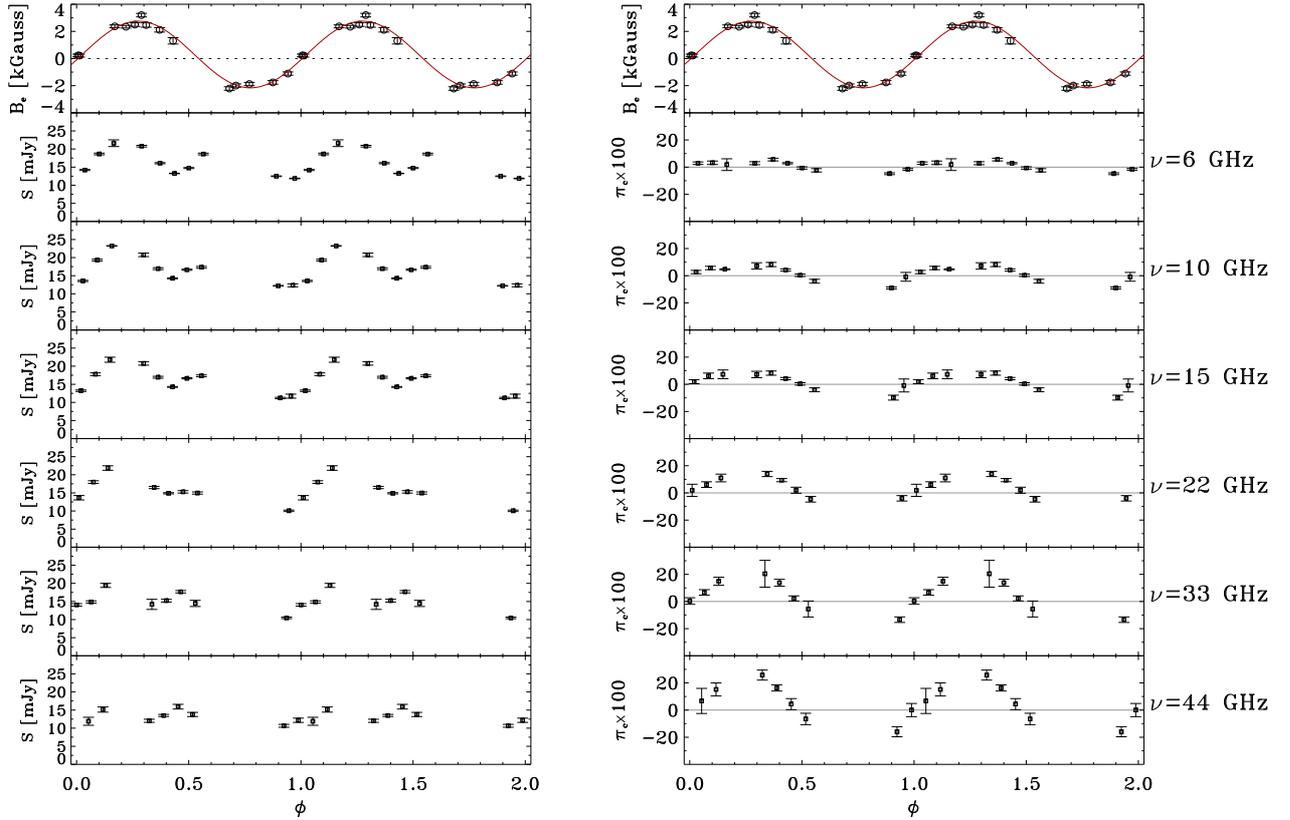}}
\caption{
Left and right top panels show the magnetic curve of HR\,7355 and data
taken by \citet{oksala_etal10} and \citet{rivinius_etal10}.  The
other left panels show the radio light curves for Stokes~I
obtained at all the observing frequencies.  Right panels 
display the rotational modulation of the fractional circularly
polarized radio emission.  
}
\label{fig_data}
\end{figure*}

\section{Observations and data reduction}
\label{sec:obs}
\subsection{Radio}

Broadband multi-frequency observations of HR\,7355 were carried out
using the Karl G.\ Jansky Very Large Array (VLA), operated by the
National Radio Astronomy Observatory\footnote{The National Radio
Astronomy Observatory is a facility of the National Science Foundation
operated under cooperative agreement by Associated Universities,
Inc.} (NRAO), in different epochs.  Table~\ref{VLA_log} reports the
instrumental and observational details for each observing epoch.
{To maximize the VLA performances
the observations were done using the full array configuration at each observing bands, 
without splitting the interferometer in sub-arrays. 
To observe all the selected sky frequencies,  
the observations were carried out cyclically varying the observing bands.}

The data were calibrated using the standard calibration pipeline,
working on the Common Astronomy Software Applications (CASA),
and imaged using the CASA task CLEAN.  Flux densities for the Stokes
I and V parameters were obtained by fitting a two-dimensional
gaussian at the source position in the cleaned maps. The size of
the gaussian profile is comparable with the array beam, indicating
that HR\,7355 is unresolved for all the analyzed radio frequencies.
The minimum array beam size is 
$0.18 \times 0.12$ [arcsec$^2$], obtained with the BnA array
configuration at 44 GHz. The errors were computed as the quadratic sum 
of the flux density error,
derived from the bidimensional gaussian fitting procedure,
and the map rms measured in a field area lacking in radio sources.

\begin{table} 
\caption{VLA observing log, Code: 15A-041} 
\label{VLA_log} 
\begin{center} 
\begin{tabular}{ccccccc} 
\hline           
\hline 
             \s $\nu$     &\s $\Delta \nu$        &\s Epoch    &\s conf.   &\s Flux cal   &\s Phase cal       \\ 
             \s [GHz]      &\s [GHz]                   &                    &               &                      &                             \\ 
\hline 
           \s 6/10/15    &\s 2  &\s 15-Apr-10    &\s B        &\s 3C286        &\s 1924$-$2914  \\ 
            \s 22/33/44 &\s 8 &\s 15-Apr-10    &\s B        &\s 3C286        &\s 1924$-$2914  \\ 
          \s 6/10/15     &\s 2  &\s 15-Jun-01    &\s BnA        &\s 3C286        &\s 1924$-$2914  \\ 
           \s 22/33/44  &\s 8 &\s 15-Jun-01    &\s BnA        &\s 3C286        &\s 1924$-$2914  \\ 

\hline   
\end{tabular}     
\end{center} 

\end{table} 

\subsection{X-ray}
\label{sec_x}
We retrieved and analyzed archival X-ray observations of HR~7355
obtained with the \XMM\ on 2012-09-25 (ObsID 0690210401, \citealp{naze_etal14}),
and lasting $\approx 2.5$\,hr.  All
three (MOS1, MOS2, and PN) European Photon Imaging Cameras (EPICs)
were operated in the standard, full-frame mode and a thick UV filter
\citep{turner_etal01,struder_etal01}.  The data were reduced using
the most recent calibration. The spectra and light-curves were extracted
using standard procedures from a region with diameter $\approx
15\arcsec$.  The background area was chosen to be nearby the star
and free
of X-ray sources.  To analyze the spectra we used the standard
spectral fitting software {\sc xspec} \citep{arnaud96}. The abundances
were set to solar values according to \citet{asplund_etal09}.  The
adopted distance to the star and interstellar reddening $E(B - V)$
are listed in Table~\ref{par_star}.

\section{The radio properties of HR\,7355}

\label{sec:radio}
\subsection{Radio light curves}
\label{light_curves}

The magnetosphere of HR\,7355 shows evidence of strong and variable
radio emission.  The VLA radio measurements were phase folded using
the ephemeris given by \citet{oksala_etal10}: \\

\noindent
$\mathrm {JD}=2\,454\,672.80+
        0.5214404 E ~~\mathrm{[days]}
$\\

\noindent
and are displayed in Fig.~\ref{fig_data}.  The left panels show the
new radio data for the Stokes I ($RCP+LCP$, respectively Right and
Left Circular Polarization state\footnote{VLA measurements of the
circular polarization state are in accordance with the IAU and
IEEE orientation/sign convention, unlike the classical physics
usage.}), with each observing radio frequency shown individually.
In the top panel of Fig.~\ref{fig_data} is also shown the variability
of the logitudinal component of the magnetic field ($B_{\rm e}$)
\citep{oksala_etal10,rivinius_etal10}.

The radio light curves for Stokes~I are variable at all observed
frequencies.  Relative to the median, the amplitudes of the variation,
with frequency are respectively: $\approx 60$\% at 6 GHz, $\approx
65$\% at 10 GHz, $\approx 62$\% at 15 GHz, $\approx 77$\% at 22
GHz, $\approx 60$\% at 33 GHz and $\approx 39$\% at 44 GHz.

The JD$_0$ of the HR\,7355 ephemeris refers to the minimum of
the photometric light curve \citep{oksala_etal10}, corresponding
to a null in the effective magnetic field curve, and to a minimum
emission for the H$\alpha$ \citep{rivinius_etal13}.  Interestingly, the radio
light curves at $\nu \leq 15$ GHz show an indication of a minimum
emission close to $\phi=0$ (see the left panels of Fig.~\ref{fig_data}).

Despite not having full coverage of rotational phase,
the radio light curve at $\nu \leq 15$\,GHz evidences a maximum
at $\phi \approx 0.2$, close to the maximum effective magnetic field
strength, followed by another minimum that becomes progressively
less deep with increasing frequency.  The rotational phases covering
the negative extrema of the magnetic curve are not observed at 
radio wavelengths, but the rising fluxes suggest that the radio
light curves at  $\nu \leq 15$ could be characterized by another
maximum.  The radio data for total intensity seems to show
2 peaks per cycle, that are related to the two
extrema of the magnetic field curve.  
Comparison between the
radio and the magnetic curves also indicates
a phase lag between the radio light curves and the magnetic one.

\begin{figure}
\resizebox{\hsize}{!}{\includegraphics{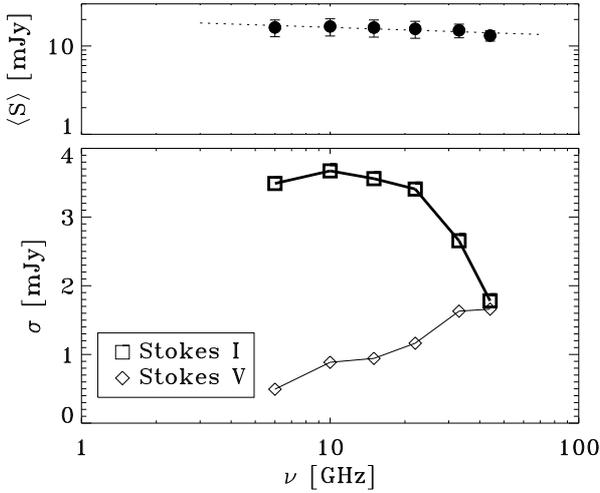}}
\caption{Top panel: radio spectrum of HR\,7355 obtained
averaging all the VLA measurements performed at the same observing band. 
The error bars are for the standard deviation of the
averaged data. The dotted line represents the power-law that best fit 
the average spectrum (spectral index $\approx 0.1$).
Bottom panel: standard deviations of the radio measurements, respectively 
for the Stokes I and V.}
\label{spe_sigma}
\end{figure}

At the higher frequencies ($\nu \geq 22$ GHz), the shapes of the
light curve are more complex, and any relation with variability
in $B_{\rm e}$
is no longer simple.  Furthermore, it appears that the
average radio spectrum of HR\,7355 is relatively flat from 6 to 44
GHz (c.f., top panel in Fig.~\ref{spe_sigma}).  The error bar of each
point shown in the figure is the standard deviation of the measurements
performed at a given frequency.  The spectral index of the mean
radio spectrum of this hot magnetic star is close to $-0.1$, like
free-free emission from an optically thin thermal plasma.  Hence, 
without any information regarding the fraction of the circularly polarized 
radio emission and its variability, the total radio 
intensity alone can easily be mistakenly attributed to 
a Bremsstrahlung radiation.
Interestingly, a flat radio spectrum has been already detected 
in some others magnetic chemically peculiar stars  \citep{leone_etal04}.

\subsection{Circular polarization}

Circularly polarized radio emission 
is detected from HR\,7355 above the
3$\sigma$ detection level, revealing
a non-thermal origin for the radio emission.  The
right panels of Fig.~\ref{fig_data} show the fraction
$\pi_{\mathrm c}$ of the circularly polarized flux density (Stokes
V / Stokes I, where Stokes V$= RCP-LCP$) as a function of the
rotational phase, and for all the observed frequencies.  In the
top right panel of Fig.~\ref{fig_data}, the magnetic field curve is
again shown for reference.

$\pi_{\mathrm c}$ is variable as HR\,7355 rotates, and the amplitude
of the variation rises as the radio frequency increases.  It appears
that the amplitude of the intensity variation is larger when the
circular polarization is smaller.  In particular $\pi_{\mathrm c}$
ranges between:  $\approx -5\%$ to 5\%  at 6 GHz, $\approx -9\%$
to 8\%  at 10 GHz,   $\approx -10\%$ to 8\%  at 15 GHz, $\approx
-5\%$ to 14\% at 22 GHz, $\approx -13\%$ to 20\%  at 33 GHz,  $\approx
-16\%$ to 26\%  at 44 GHz.

To parameterize the amplitude of the radio light curves for the
total and polarized intensity, the bottom panel of Fig.~\ref{spe_sigma}
shows the variation of the standard deviation ($\sigma$) of all
the measurements occuring at the same frequency, as a function of 
frequency.  The standard deviation of the Stokes I measurements is
largest at $\nu\leq 22$ GHz, whereas at 33 and 44 GHz, the standard
deviations dramatically decrease, confirming the decreasing of the
light curve amplitudes discussed in Sec.~\ref{light_curves}.  By
contrast the standard deviation of the measurements of the circularly
polarized flux density increases as the frequency increases.  Considering
Fig.~\ref{spe_sigma} (bottom panel), $\sigma$ values
for the Stokes I and V measurements are evidently inversely related.

\begin{table*}
\caption[ ]{}
\label{mod_par}
\begin{tabular}{lccl}
\hline
\hline
\multicolumn{4}{l}{Free parameters}\\
\hline
   &Symbol         &Range          &Simulation step\\
\hline
Alfv\'en radius [R$_{\ast}$]                                &$R_\mathrm{A}$      &$8$ -- $50$            &$\Delta \log R_\mathrm{A}\approx0.1$ \\
Thickness of the middle magnetosphere [R$_{\ast}$]                     &$l$                 &$1$ -- $40$          &$\Delta \log l\approx0.2$ \\
Non thermal electron density [cm$^{-3}$]                              &$n_\mathrm{r}$      &$10^2$ -- $10^{5}$          &$\Delta \log n_\mathrm{r}\approx 0.1$  \\
Relativistic electron energy power-law index                           &$\delta$            &2 -- 3                 &$\Delta \delta=0.5$ \\
Thermal electron density at the stellar surface [cm$^{-3}$]  &$n_\mathrm{0}$    &$10^{8}$ -- $10^{10}$  &$\Delta \log n_\mathrm{p_0}\approx0.25$ \\
\hline
\hline
\multicolumn{4}{c}{Model solutions with $\delta=2.5$}\\
\hline

Thermal electron density at the stellar surface  &\multicolumn{3}{c}{$n_\mathrm{0}=3 \times 10^9$ [cm$^{-3}$]}\\
Alfv\'en radius &\multicolumn{3}{c}{$R_\mathrm{A}=12.5$ -- $40$ [R$_{\ast}$]}\\
Column density of the non thermal electrons &\multicolumn{3}{c}{$n_\mathrm{r} \times l = 10^{(12.95\pm0.09)} \times R_\mathrm{A}^{(2.68\pm0.07)}$ [cm$^{-2}$]}\\

\hline
\hline
\multicolumn{4}{c}{Model solutions with $\delta=2$}\\
\hline

Thermal electron density at the stellar surface  &\multicolumn{3}{c}{$n_\mathrm{0}=2$ -- $5 \times 10^9$ [cm$^{-3}$]}\\
Alfv\'en radius &\multicolumn{3}{c}{$R_\mathrm{A}=10$ -- $40$ [R$_{\ast}$]}\\
Column density of the non thermal electrons &\multicolumn{3}{c}{$n_\mathrm{r} \times l = 10^{(12.48\pm0.04)} \times R_\mathrm{A}^{(2.44\pm0.03)}$ [cm$^{-2}$]}\\

\hline
\end{tabular}
\end{table*}

Comparing the curves of variation of $\pi_{\mathrm c}$ with the
magnetic field curve, a positive degree of circular
polarization is detected when the north magnetic pole is close to
the line-of-sight, and is negative when the south pole is most
nearly aligned with the viewing sightline.  When
the magnetic poles are close to the direction of the line-of-sight,
we observe most of the radially oriented field lines.  In this case
the gyrosynchrotron mechanism gives rise to radio emission that is partially
polarized, respectively right-handed for the north pole and left-handed
for the south pole.  This behavior of the gyrosynchrotron polarized
emission has already been recognized, at $\nu \leq 15$ GHz, in the
cases of CU\,Vir \citep{leto_etal06} and $\sigma$\,Ori\,E
\citep{leto_etal12}, that have magnetospheres defined  mainly by a
magnetic dipole, similar to the case of HR\,7355.  Our new radio
measurements show, for the first time, this 
behavior persisting up to $\nu=44$ GHz for HR\,7355.
Furthermore, the light curves of $\pi_{\mathrm c}$ show
that the magnetic field component close to the stellar
surface can be traced with the circularly polarized emission at high
frequency.

\begin{figure}
\resizebox{\hsize}{!}{\includegraphics{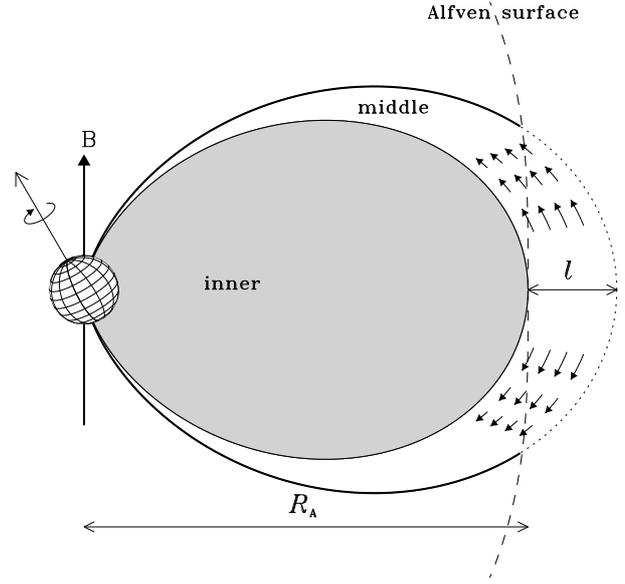}}
\caption{Meridional cross-section of the magnetospheric model for a
hot magnetic star, characterized by a simple dipolar magnetic field,
and with the dipolar axis misaligned with respect to the rotation axis
(the misalignment amplitude is arbitrary).  The grey area indicates
the thermal plasma trapped within the inner magnetosphere. The
equatorial region of the magnetosphere outside the Alfv\'{e}n surface
(dashed line) is a site of magnetic reconnection events, capable of
accelerating the {\it in situ} plasma.  The length ($l$) of the non-thermal
acceleration region, just outside the equatorial Alfv\'{e}n radius
($R_{\mathrm A}$), is also shown.  The magnetic shell,
middle-magnetosphere, where the non-thermal electrons (indicated
by the small vectors) propagate toward the stellar surface, radiate
by the gyro-synchrotron emission mechanism, is delimited by the two
pictured magnetic field lines.}
\label{modello}
\end{figure}

\section{The model}
\label{model}

In previous papers \citep{trigilio_etal04,leto_etal06}, we developed
a 3D model to simulate the gyrosynchrotron radio emission
arising from a stellar magnetosphere defined by a dipole.  In the
case of the hot magnetic stars, the scenario attributes the origin of their
radio emission as the interaction between the large-scale
dipolar magnetic field and the radiatively driven stellar wind.

Following this model the plasma wind progressively accumulates in
the magnetospheric region where the magnetic field lines are closed
(inner magnetosphere).  The plasma temperature linearly increases
outward, whereas its density linearly decreases
\citep{babel_montmerle97,ud-doula_etal14}.  
Outside the Alfv\'{e}n surface, the magnetic tension is not able to force co-rotation of the plasma.
Similar to the case of Jupiter's magnetosphere \citep{nichols11},
the co-rotation breakdown powers a current sheet system
where magnetic reconnection accelerates the local plasma
up to relativistic energies \citep{usov_melrose92}.
A fraction of the non-thermal electrons, assumed to have a power-law energy
spectrum and an isotropic pitch angle distribution (i.e., the angle between the
directions of the electron velocity and the local magnetic field
vector), can diffuse back to the star within a magnetic
shell that we here designate as the ''middle magnetosphere''.  This
non-thermal electron population has a homogeneous spatial density
distribution within the middle magnetosphere, owing to magnetic
mirroring.  A cross-section of the stellar magnetosphere model is
pictured in Fig.~\ref{modello}.

The non-thermal electrons moving within the middle magnetosphere
radiate at radio wavelengths by the gyro-synchrotron emission
mechanism.  To simulate the radio emission arising from these
non-thermal electrons, the magnetosphere of the
star is sampled in a three dimensional grid, and the physical
parameters needed to compute the gyrosynchrotron emission and
absorption coefficients are calculated at each grid point.

As a first step, we set the stellar geometry: rotation
axis inclination ($i$), and tilt of the dipole magnetic axis ($\beta$),
and the polar field strength ($B_{\mathrm p}$).  In the stellar
reference frame, assumed with the $z$-axis coinciding with the
magnetic dipole axis, the space surrounding the star is sampled
in a 3D cartesian grid, and the dipolar magnetic field vector
components are calculated at each grid point.  Given the
stellar rotational phase ($\phi$), the field topology is 
then rotated in the observer reference frame (see procedure described
in App.~A of \citealp{trigilio_etal04}).

In the second step, we locate the magnetospheric subvolume where the unstable
electron population propagates.  This spatial region is delimited by
two magnetic field lines.  The inner line intercepts the magnetic
equatorial plane at a distance equal to the Alfv\'{e}n radius
($R_{\mathrm A}$).  The outer line intercepts the equatorial
plane at a distance $R_{\mathrm A}+l$, with $l$ being the width of the current
sheet where magnetic reconnection accelerates the local plasma
up to relativistic energies.  Within each grid point of the middle
magnetosphere, the non-thermal electrons have a constant number
density ($n_{\mathrm r}$).  By contrast the inner magnetosphere is 
filled by a thermal plasma with density and temperature that are
functions of the stellar distance as previously described.

In the third step, given the observing radio frequency $\nu$, we
calculate the emission and absorption coefficients for the
gyro-synchrotron emission \citep{ramaty69} at the grid points that
fall within the middle magnetosphere.  For each grid point of the
inner magnetosphere, the free-free absorption
coefficient \citep{dulk85}, the refractive index, and the polarization
coefficient for the two magneto-ionic modes \citep{klein_trotter84}
are computed.
We are able to solve numerically the radiative transfer equation
along the directions parallel to the line-of-sight for the Stokes
I and V (as described in the App.~A of \citealp{leto_etal06}).
Scaling the result for the stellar distance, and repeating these
operations as a function of the rotational phase, $\phi$, 
synthetic stellar radio light curves are calculated, and
then simulations are compared with observations.

\begin{figure*}
\resizebox{\hsize}{!}{\includegraphics{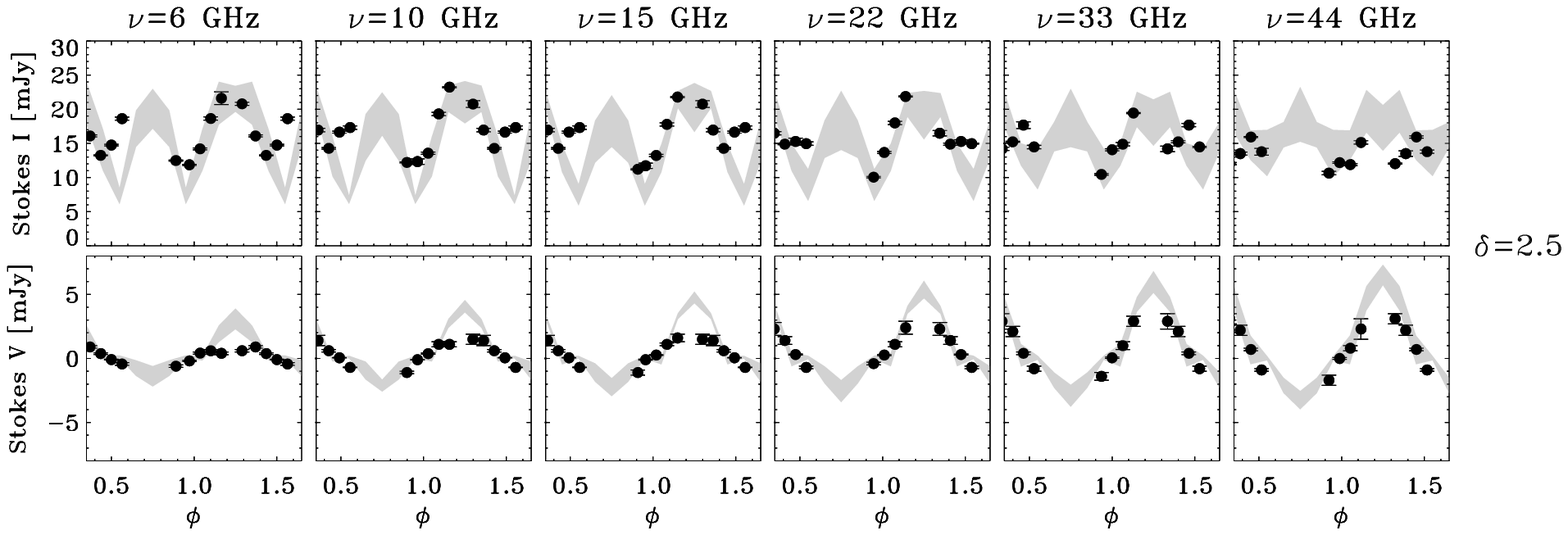}}
\resizebox{\hsize}{!}{\includegraphics{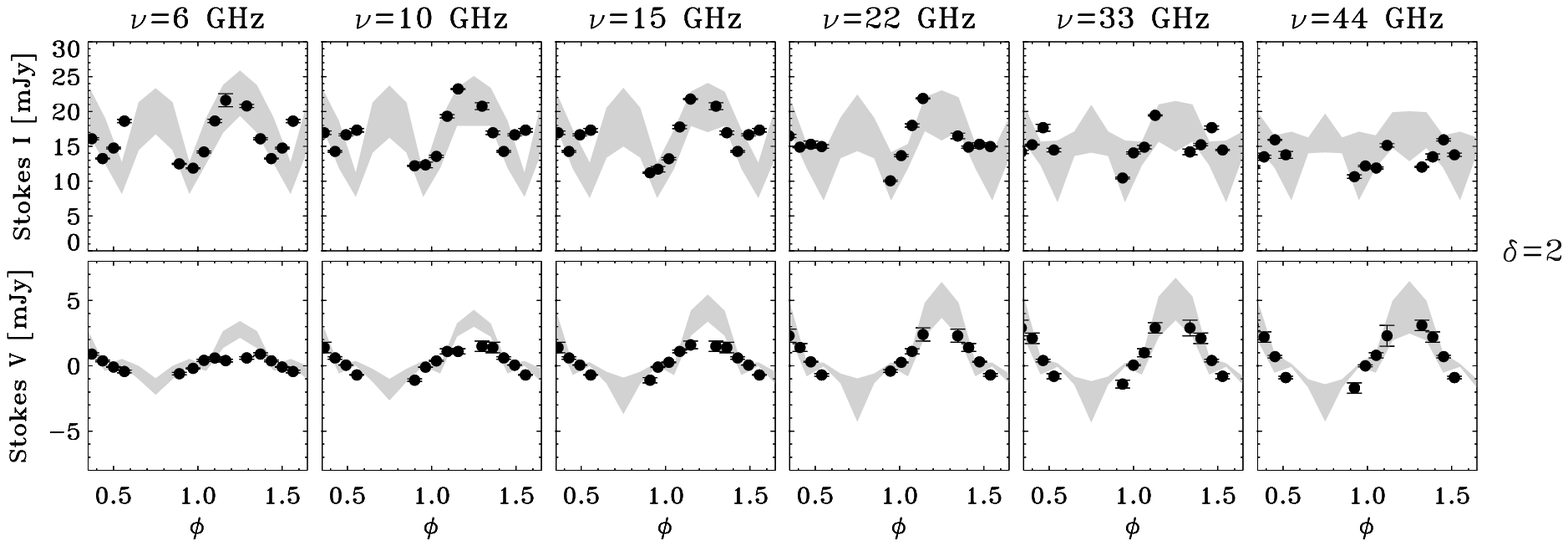}}
\caption{The filled circles are the HR\,7355 observed radio 
light curves, respectively of the total intensity (Stokes I), 
and of the circularly polarized flux density (Stokes V).
The grey areas are the envelope of the modeled light curves 
{obtained by using the combinations of the free parameters able to generate synthetic light curves that close match the observed ones}. 
The corresponding model solutions are listed in Table~\ref{mod_par}.
Top panels shown the model solutions with $\delta=2.5$, bottom panels those with $\delta=2$.}
\label{hr7355_radio_simul}
\end{figure*}

\subsection{Modeling the HR\,7355 radio emission}

On the basis of the model described in previous section, we seek to reproduce
the multi-wavelength radio light curves of HR\,7355
for Stokes I and V.  The already known stellar parameters of HR\,7355, 
needed for the simulations are listed in Table~\ref{par_star}.
The Alfv\'{e}n radius ($R_{\mathrm A}$) 
and the length of the current sheet ($l$) have been assumed as free
parameters.  For the sampling step, we adopt a variable grid with a
narrow spacing (0.1 R$_{\ast}$) for distances lower then 8 R$_{\ast}$,
a middle spacing (0.3 R$_{\ast}$) between 8 and 12
R$_{\ast}$, and a rough spacing  (1 R$_{\ast}$) for
distances beyond 12 R$_{\ast}$.

Following results obtained from simulations of radio
emissions of other hot magnetic stars \citep{trigilio_etal04,leto_etal06},
the low-energy cutoff of the power-law electron energy
distribution has been fixed at 100 keV, corresponding to a Lorentz factor $\gamma = 1.2$.
The temperature of the thermal plasma at the stellar surface has
been set equal to the photospheric one (given in Tab.~\ref{par_star}),
whereas its density ($n_{0}$) has been assumed as a free parameter.
The assumed values of the model, free parameters, and the corresponding
simulation steps are listed in Table~\ref{mod_par}. 
Adopting these
stellar parameters, we were able to simulate radio light curves for
the Stokes I and V that closely resemble the measurement of HR\,7355.
The corresponding ranges of the model parameters are reported in Table~\ref{mod_par}.

{The Fig.~\ref{hr7355_radio_simul} displays the envelope of the simulated light curves, 
for the Stokes I and V respectively, that closely match the observed ones.
This envelope was obtained from the simultaneous  visualization of the
whole set of simulations performed using the combinations of the model free parameters
listed as model solutions in Table~\ref{mod_par}.}
The simulations indicate that
gyro-synchrotron emission from a dipole-shaped magnetosphere can closely
reproduce the observations of HR\,7355.  The low-frequency Stokes I 
radio emission shows a clear phase modulation, that
becomes progressively less evident as the frequency increases.
Conversely the simulations of the light curves for the Stokes V
indicates that the circularly polarized emission is strongly
rotationally modulated, with an amplitude that increases with frequency.
Such behavior of the simulated
radio light curves is consistent with the measurements.

\begin{figure}
\resizebox{\hsize}{!}{\includegraphics{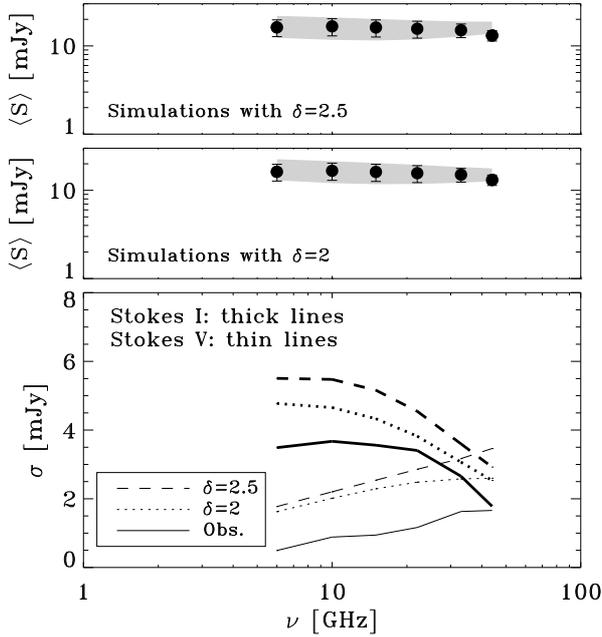}}
\caption{Top and middle panels: like Fig.~\ref{spe_sigma}, 
the filled circles are the average measured radio spectrum
of HR\,7355. The grey areas are the envelope of the averages from simulated 
radio spectra corresponding to the model solutions
with $\delta=2.5$ (top panel), and with $\delta=2$ (middle panel). In the 
bottom panel, the 
standard deviations of the observations (continuous lines) and simulations 
(dashed lines corresponding to $\delta=2.5$,
and dotted lines corresponding to $\delta=2$) are compared. The thick lines 
correspond to the total intensity (Stokes I),
the thin lines to the circularly polarized flux density (Stokes V).}
\label{spe_sigma_simul}
\end{figure}

{To highlight the close match between simulations and observations,
we also compared the simulated radio spectra with the observed spectrum.
The synthetic spectra were realized
averaging the simulated light curves at each frequency.} 
In the top and middle panels of Fig.~\ref{spe_sigma_simul},
the observed spectrum of HR\,7355 is shown again, and the superimposed shaded
area represents the 
{envelope of the simulated spectra for the Stokes I.
In the bottom panel of Fig.~\ref{spe_sigma_simul}
the standard deviation ($\sigma$)
of the observed and simulated multi-frequency light curves (Stokes I and V) have been compared.
In the case of the model simulations, more than one spectrum was
produced. The $\sigma$ values pictured in the bottom panel of Fig.~\ref{spe_sigma_simul}
are the averages of the standard deviations corresponding to the
whole set of simulated light curves.}
%
The top panel of Fig.~\ref{spe_sigma_simul} refers to 
the model simulations with parameters $\delta=2.5$,
whereas the middle panel is for the $\delta=2$ case.
The $\sigma$ of the simulated
HR\,7355 radio emission seems to be larger than the observed ones.
Such behavior is confirmed when looking at the bottom panel of
Fig.~\ref{spe_sigma_simul}.  
In particular the $\sigma$ of the light curves with $\delta=2.5$
are highest. This behavior suggests that the value of $\delta=2$,
for the spectral index of the non-thermal electron, could be close to the true value.
But we must also take into account that,
the magnetosphere of this rapidly rotating star could be oblate, 
whereas our model assumes a simple dipole.
The stretching of the magnetosphere could affect the magnetic field topology of the regions
where the radio emission at the observed frequencies originate.
The effect of the plasma inertia to the magnetic field configuration is an issue
outside the limit of our model. 
In any case, this mismatch between the dipolar and the true stellar magnetic topology
could explain the differences between observations and simulations.
Furthermore, has been proven that within the HR\,7355 magnetosphere
there are cloud the dense plasma (with linear size $\approx 2$ R$_{\ast}$)
co-rotating with the star \citep{rivinius_etal13}, that could affect the 
rotational modulation of the stellar radio emission.
The modeling approach followed to simulate the radio light curves of HR\,7355
does not take into account for the presence of such material.
On the basis of this considerations we cannot exclude that the spectral
index of the non thermal electrons could be close to $\delta=2.5$.
In any case, the higher dispersion of the simulations
with respect to the observations can be explained as a consequence of
the coverage for the observed radio light curves not being complete.  
In fact, we are missing some portions of the light curves that are expected
to be highly variable.  On the other hand, the frequency dependence of the
standard deviations of the simulated Stokes I and V radio light
curves are similar to the observed ones (see bottom panel of
Fig.~\ref{spe_sigma_simul}).  This is further evidence of the
good fit of our model for describing the radio magnetosphere of
HR\,7355.

\section{The magnetosphere of HR\,7355}
\label{sec:magnet}
\subsection{Radio diagnostic}
\label{sec:radiodiagnostic}

\begin{figure}
\resizebox{\hsize}{!}{\includegraphics{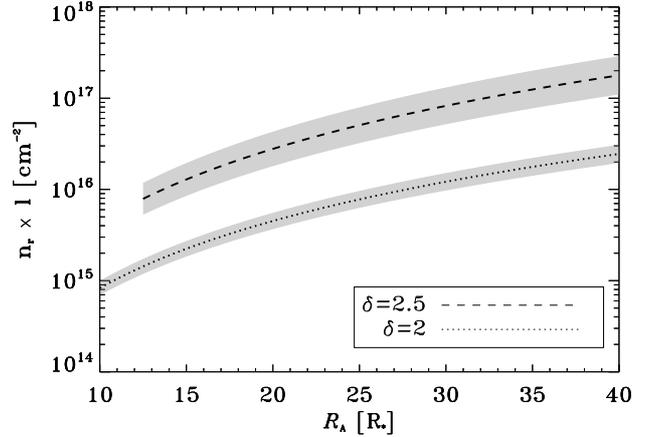}}
\caption{Graphical view of the analytic equation that describes the column density of the non-thermal electrons (given in Table~\ref{mod_par}),
calculated close to the acceleration site, as a function of the values of $R_{\mathrm A}$ that are
able to give simulated radio light curves matching the observed ones.
The dotted line corresponds to the model solutions with $\delta=2$; the dashed line with $\delta=2.5$.
Note acceptable model solutions are not found
for $R_{\mathrm A} < 12.5$ R$_{\ast}$  with $\delta=2.5$. The grey areas
highlight the solution uncertainty.}
\label{col_dens}
\end{figure}

Analysis of model solutions for
the observed multi-wavelength radio light curves of HR\,7355,
respectively for the Stokes I and V, can be used to 
constrain the physical conditions of the magnetosphere of this
hot  star.  The thermal electron density at the stellar
surface ($n_{0}$) is well constrained.  We found acceptable
light curves for Alfv\'{e}n radii greater than 10 R$_{\ast}$.  The
other two model free parameters are degenerate, namely
the non-thermal electron density ($n_{\mathrm r}$) and the length
of the current sheet ($l$). The product of these two parameters is
the column density of relativistic electrons at the Alfv\'{e}n
radius. We found that the column density is a function of $R_{\mathrm
A}$, the mathematical relationship, obtained by fitting these
parameters, is provided in Table~\ref{mod_par}, and 
pictured in Fig.~\ref{col_dens}.

\begin{figure}
\resizebox{\hsize}{!}{\includegraphics{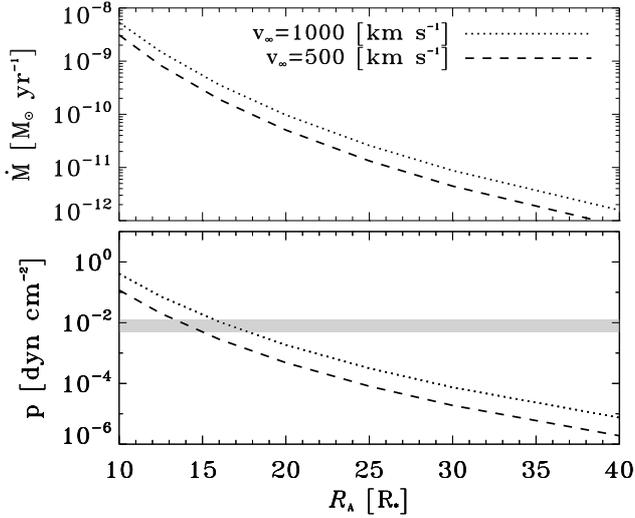}}
\caption{Top panel: values of the wind mass-loss rate corresponding
to the range of Alfv\'{e}n radii derived by the model simulations
of the multi-wavelength radio light curves of HR\,7355. The values
of $\dot{M}$ have been derived assuming two reasonable values for
a main sequence B-type star wind terminal velocity.  Bottom
panel: wind pressure at the Alfv\'{e}n radius. Dotted line 
refers to the case $v_{\infty}=500$ [km s$^{-1}$], dashed line to
$v_{\infty}=1000$ [km s$^{-1}$]. The grey area indicates the
range of pressures from the thermal plasma trapped within the
inner magnetosphere. }
\label{pterm_ram}
\end{figure}

The value of the Alfv\'{e}n radius 
is related to the wind of HR\,7355.  In
\citet{trigilio_etal04} we computed $R_{\mathrm A}$ given the magnetic
field strength, the wind mass-loss rate, its terminal velocity
($v_{\infty}$), the stellar radius, and the rotation period.  In
the present analysis, we reverse this approach: given $v_{\infty}$
and the rotation period, we estimate the mass-loss rates ($\dot{M}$)
of HR\,7355 that are compatible with the values of $R_{\mathrm A}$
listed in Table~\ref{mod_par}.  We assume two values of wind terminal
velocity that are reasonable for a main sequence B type star
\citep{prinja89,oskinova_etal11,kritcka14}: $v_{\infty} =500$ and
1000 [km s$^{-1}$].  Fig.~\ref{pterm_ram} shows the values of
$\dot{M}$, and the corresponding pressure, as a function of $R_{\mathrm
A}$.  The highest values of $R_{\mathrm A}$ need a low
wind mass-loss rate.

The model simulation provide an
estimate for the density of the thermal plasma trapped within the inner
magnetosphere of HR\,7355.  The adopted radial dependence for the
plasma temperature and density are respectively: $n=n_\mathrm{0}
r^{-1}$ and $T=T_\mathrm{eff} r$, hence the thermal pressure
($p=k_{\mathrm B} n T$) is constant inside the inner magnetosphere.
In steady state $p=p_{\mathrm {ram}}$, where $p_{\mathrm {ram}}$
is the wind ram pressure.  In the bottom panel of Fig.~\ref{pterm_ram}
the grey area represents the thermal pressure of the plasma trapped
in the inner magnetosphere.  Those solutions that do not satisfy
the above equality condition cannot be considered valid.  The average
Alv\'{e}n radii that are physically plausible are listed in
Table~\ref{tab_wind}.  The corresponding wind mass loss rate, 
the density of the wind at the Alv\'{e}n radius, the average 
thermal temperature of the plasma trapped within the inner 
magnetosphere, as well as the corresponding emission measure 
are also listed in Table~\ref{tab_wind}.

In the case of a dipolar shaped magnetosphere (see Fig.~\ref{modello}),
the radiatively driven stellar wind can freely propagate only from
the northern and southern polar caps. As a
consequence, the actual mass loss rate ($\dot{M}_{\mathrm {act}}$)
is a fraction of $\dot{M}$. The fraction of the wind that freely propagates 
can be estimated from the ratio between the two polar caps area and the whole surface.
The polar caps area is derived from the relation
defining the dipolar magnetic field line: $r=R_{\mathrm A} \cos \lambda ^2$
(where $\lambda$ is the magnetic latitude). In fact,
the point where the field line, with a given $R_{\mathrm A}$, crosses the stellar surface
individuates the latitude of the polar cap.
The values of $\dot{M}_{\mathrm {act}}$, listed in Table~\ref{tab_wind},
are in good agreement with those obtained from the UV spectral 
analyses of HR\,7355 (see Sec.\,\ref{sec:uv}) and other B-type stars with 
similar spectral types \citep{prinja89,oskinova_etal11,kritcka14}

The indirect  evaluation of the linear extension of the radio emitting region
is also useful for estimating the brightness temperature of HR\,7355.
The average flux densities for HR\,7355 are $\approx 15.5$ mJy, in the frequency range 6--44 GHz.
Assuming the mean equatorial diameter of the Alfv\'{e}n surface (31 R$_{\ast}$) for the source size,
the corresponding brightness temperature is $T_{\mathrm {bril}} \approx 3 \times 10^{10}$ [K].
The above estimate reenforces the conclusion that
the radio emission from HR\,7355 has a non-thermal origin.

It is instructive to compare  the results obtained from the 
analysis of radio emission from HR\,7355 and from the Ap 
star CU\,Vir conducted using similar 
approach \citep{leto_etal06}. 
For example, the wind electron density number at the Alv\'{e}n surface,
$n_{\mathrm w}(R_\mathrm{A})$, has similar values for both stars. 
The estimated column density of the relativistic electrons at the Alv\'{e}n radius lies
in the range 3.2--$4.6 \times 10^{14}$ [cm$^{-2}$] in the case of CU\,Vir, 
versus a column density that ranges between $1.9 \times10^{15}$ and $3.0\times10^{15}$ [cm$^{-2}$] for HR\,7355.
This is even higher in the case of the HR\,7355 model solutions with $\delta=2.5$:
the derived range is 1.1--$1.8\times10^{16}$ [cm$^{-2}$].
For the case of $\delta=2$, column density of  the non-thermal electrons for HR\,7355 
is higher by about an order of magnitude as compared to CU~Vir.

\begin{table}
\begin{center}
\caption[ ]{Derived parameters of HR\,7355}
\label{tab_wind}
\footnotesize
\begin{tabular}[]{l cc c cc}
\hline
\hline
                           &$v_{\infty}$                   &$<R_\mathrm{A}>$              &         &$v_{\infty}$                   &$<R_\mathrm{A}>$   \\
                                                   \cline{2-3}                                                                              
                                                                                                                                                         \cline{5-6}      
                           &\s[km s$^{-1}$]             &\s[R$_{\ast}$]                        &         &\s[km s$^{-1}$]             &\s[R$_{\ast}$]                 \\
                                                   \cline{2-3}                                                                              
                                                                                                                                                         \cline{5-6}                                       
                           &500                                &14                                           &          &1000                             &17                                 \\             
\hline
                                                                                                                                                                      
$<\dot{M}>$ \s[M$_{\odot}$ yr$^{-1}$]                              &\multicolumn{2}{c}{$4.2\times10^{-10}$}            &          &\multicolumn{2}{c}{$2.5\times10^{-10}$}     \\                                                                                                                                                                     
$\dot{M}_{\mathrm {act}}$ \s[M$_{\odot}$ yr$^{-1}$]        &\multicolumn{2}{c}{$1.5\times10^{-11}$}            &           &\multicolumn{2}{c}{$0.7\times10^{-11}$}    \\               
$n_{\mathrm w}(R_\mathrm{A})$ \s[cm$^{-3}$]              &\multicolumn{2}{c}{$2.0\times 10^{6}$}               &          &\multicolumn{2}{c}{$0.4\times 10^{6}$}     \\  
$<T>$ \s[MK]                                                                          &\multicolumn{2}{c}{$0.16$}                           &          &\multicolumn{2}{c}{$0.19$}                   \\ 
$EM$ \s[$10^{55}$ cm$^{-3}$]                                           &\multicolumn{2}{c}{$2.22$}                           &          &\multicolumn{2}{c}{$2.72$}     \\ 
\hline
\end{tabular}
\end{center}
\end{table}

Under the reasonable assumption that HR\,7355 and CU\,Vir have similar non-thermal
acceleration efficiencies, 
the higher non-thermal electron column density of HR\,7355 could be explained 
if it is characterized by a more extended acceleration region as compared to CU\,Vir.  
The magnetosphere of HR\,7355 is bigger than CU\,Vir,
with $R_{\ast}=3.69$ versus 2.06 R$_{\odot}$ for CU\,Vir
\citep{kochukhov_etal14}, and so the linear size of the acceleration
region ($l$) will be consequently wider for HR\,7355.
Furthermore, the B2~type star HR\,7355 has a higher stellar mass compared to the
Ap~star CU\,Vir, 6 versus 3.06 M$_{\odot}$ \citep{kochukhov_etal14}.
Their Kepler corotation radii ($R_{\mathrm K}=(G M_{\ast} / \omega^2)^{1/3}$)
are respectively: 1.3 R$_{\ast}$ for HR\,7355, 1.9  R$_{\ast}$ for CU\,Vir.
Comparing the above estimated values of $R_{\mathrm K}$ with 
the average Alfv\'{e}n radii, respectively 15.5  for HR\,7355, 
and 14.5 [R$_{\ast}$]  for CU\,Vir.
The ratio $R_{\mathrm A}/R_{\mathrm K}$ for the HR\,7355 CM magnetosphere is $\approx 11.9$,
versus $\approx 7.6$ in the case of CU\,Vir.
The above estimation highlights that HR\,7355 is characterized by a larger 
magnetospheric volume maintained in rigid co-rotation compared with that of CU\,Vir.

\begin{figure}
\resizebox{\hsize}{!}{\includegraphics{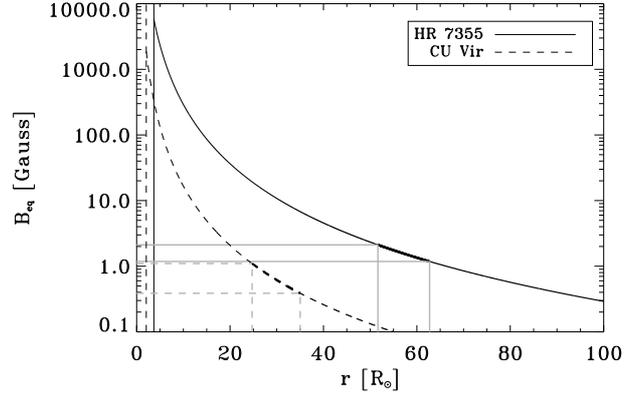}}
\caption{Radial dependence of the magnetic field strength in the equatorial plane of HR\,7355 (continuous line), and
CU\,Vir (dashed line). The ranges of the Alfv\'{e}n radii of the two stars and the corresponding magnetic field strengths are
indicated.
}
\label{b_r}
\end{figure}

We also compare the
magnetic field strength at the Alfv\'{e}n radius for both stars.
The two stars have similar rotation periods ($\approx 0.52$ d), but
HR\,7355 has a larger size, and a stronger polar magnetic field
strength (11.6~kG versus 3.8~kG, \citealp{kochukhov_etal14}).
The radial dependence for a simple magnetic dipole at the equatorial
plane is described by $B_{\mathrm {eq}}=1/2
B_{\mathrm p} (R_{\ast}/r)^3$ [Gauss], and is plotted in Fig.~\ref{b_r}.
The ranges of the allowed $R_{\mathrm A}$ values, given in units of
solar radii, are shown for both stars.  The corresponding magnetic
field strengths are derived.  Fig.~\ref{b_r} makes clear
that the current sheet region of HR\,7355 is characterized by a
magnetic field strength of roughly double the value of the case for
CU\,Vir. From a purely qualitative point-of-view, it is reasonable
to assume the non-thermal acceleration process operates within a
thicker middle magnetosphere for HR\,7355.

At the distances of the two analyzed stars, $D=236$ pc for HR\,7355
and $D=80$ pc for CU\,Vir, their radio luminosities are
respectively, $\approx 10^{18}$ [erg s$^{-1}$ Hz$^{-1}$],
obtained using the average radio flux density measured in this
paper; and $\approx 3\times 10^{16}$ [erg s$^{-1}$ Hz$^{-1}$], 
using the mean of the measured flux densities listed in
\citet{leto_etal06}.  As discussed above, the two stars are
characterized by different radio emitting volumes. For HR\,7355 the
non-thermal electrons also travel within magnetospheric regions at
higher magnetic field strength.  Using a model for gyrosynchrotron
emission, the magnetic field strength directly affects the observed
radio flux density level \citep{leto_etal06}.  
Taking into account
the various physical differences, we are able to explain qualitatively why
HR\,7355 is a brighter radio source as compared to CU\,Vir.

\subsection{X-ray diagnostic}
\label{sec_xray}

%

The X-ray flux of HR\,7355 in the 0.2--10\,keV band  
measured by the \XMM\ is $\approx 1.6 \times 10^{-13}$\,erg cm$^{-2}$
s$^{-1}$ (see Table\,\ref{par_xray}). This is orders of 
magnitude higher that may be expected if plasma emitting in radio regime would 
be solely responsible for the X-ray generation --
using the average temperature and emission measure listed in Table~\ref{tab_wind}, 
the expected X-ray flux is only $\approx 10^{-15}$ erg cm$^{-2}$ s$^{-1}$.  
Thus, a cold thermal plasma component responsible for the radio emission alone 
cannot explain the observed X-rays from HR\,7355.

\begin{table}
\begin{center}
\caption[ ]{The X-ray spectral parameters derived from the \XMM\ EPIC 
observations of
HR\,7355 assuming two-temperature CIE plasma (apec) (tbabs) model and 
apec+powerlaw model, both models corrected for the interstellar absorption. The 
values which have no error have been frozen during the fitting process. 
{The spectral fits corresponding to the the apec+powerlaw model are shown in Fig.~\ref{fig_xray}.}
}
\label{par_xray}
\footnotesize
\begin{tabular}[]{lc}
\hline
\hline
$N_{\mathrm H}^{\mathrm a}$ [$10^{20}$ cm$^{-2}$]                                               &3.2                    \\
\hline
\multicolumn{2}{l}{Two temperature thermal model}\\
\hline

$kT_1$ [keV]                                                                                                                      &$0.9\pm0.2$   \\
$EM_1$ [$10^{51}$ cm$^{-3}$]                                                                                     &$4.4\pm1.9$   \\
$kT_2$ [keV]                                                                                                                      &$3.9\pm0.8$   \\
$EM_2$ [$10^{51}$ cm$^{-3}$]                                                                                     &$46.2\pm4.9$   \\
$\langle k T \rangle \equiv \sum_i k T_i \cdot EM_i / \sum_i EM_i$ [keV]               &3.6   \\
Flux$^b$ [$10^{-13}$ erg cm$^{-2}$ s$^{-1}$]                                                            &1.6   \\
\hline

\multicolumn{2}{l}{Thermal plus power-law ($A(E)=KE^{-\alpha}$) model}\\

\hline

$kT_1$ [keV]                                                                                                                      &$1.0\pm0.1$   \\
$EM_1$ [$10^{51}$ cm$^{-3}$]                                                                                     &$6.5\pm1.5$   \\
$\alpha$                                                                                                                             &$1.7\pm0.1$   \\
$K$ [photons keV$^{-1}$ cm$^{-2}$ s$^{-1}$ at 1 keV]                                            &$(1.9\pm0.2)\times10^{-5}$   \\
Flux$^b$ [$10^{-13}$ erg cm$^{-2}$ s$^{-1}$]                                                            &1.7   \\

\hline

$L_{\mathrm X}^{\mathrm b}$ [erg s$^{-1}$]                                                                &$1.1 \times 10^{30}$   \\
$\log L_{\mathrm X} / L_{\mathrm {bol}}$                                                                      &$-6.5$   \\
$L_{\mathrm X}/L_{\mathrm {\nu,rad}}$ [Hz]                                                                &$1.1 \times 10^{12}$  \\

\hline

\end{tabular}
\begin{list}{}{}
\item[$^{\mathrm{a}}$] correspond to the ISM hydrogen column density
\item[$^{\mathrm{b}}$] dereddened; in the 0.2--10 keV band
\end{list}
\end{center}
\end{table}

Comparing the X-ray and the radio emission of HR\,7355 to that of 
late-type stars reveals significant differences
(see Table~\ref{par_xray}). HR\,7355 violates
the empirical relation coupling the X-ray and radio luminosities 
of magnetically active stars 
($L_{\mathrm X} / L_{\nu,\mathrm {rad}} \approx 10^{15.5}$ Hz, 
\citealp{guedel_benz93,benz_guedel94}),
which is valid among stars distributed within a wide range of 
spectral classes (from F to early M stars).
This is clear evidence that the physical mechanisms for the radio and X-ray
emissions operating in an early B-star like HR\,7355 are 
distinct from coronal mechanisms operating in
the intermediate- and low-mass main sequence stars.
Yet, somewhat surprisingly, the deviation of the early type
HR\,7355  from the Gu\"{e}del--Benz relation is similar to that for the stars
at the bottom of the main sequence -- the ultra cool dwarfs with 
spectral type later than M7
\citep*{berger_etal10,williams_etal14,lynch_etal16}.
These important similarities  between 
active ultra cool dwarfs and a strongly magnetic B star 
indicate that radio and X-ray emission in their magnetospheres 
may be produced by related  physical mechanisms and provide 
useful hints for the latter.

According to the MCWS model, the thermal plasma responsible for 
X-ray emissions from magnetic B-type stars is produced by stellar
wind streams colliding at the magnetic equator. 
The radio wavelengths 
are instead sensitive
to only the cold thermal plasma that accumulates at the higher magnetic
latitudes.
Consequently, the X-ray emission provides a different set of
constraints on
the physical conditions in the magnetosphere of
HR\,7355.

We have analyzed the archival \XMM\ measurements. First, 
the observed spectrum in 0.2-10.0\,keV band was fit with a thermal 
two-temperature
spectral model that assumes optically thin plasma in collisional
equilibrium.  The fit is statistically significant, with reduced
$\chi^2 =0.72$ for 88 degrees of freedom.  The model fit parameters
are shown in Table~\ref{par_xray}.  The two-temperature components
are well in accord with the values listed by \citet{naze_etal14}.
The thermal plasma is extraordinarily hot, with the bulk of the plasma
at a temperature 3.6\,keV (40\,MK).  This is significantly hotter
than usually found in magnetic B-stars 
\citep{oskinova_etal11,naze_etal14,ignace_etal13,oskinova_etal14}.  

In the framework of the MCWS model, the wind plasma streams that
collide at the magnetic equator give rise to a shock that heats the
plasma. Hence, the maximum temperature follows from a Rankine-Hugoniot 
condition and cannot exceed a value determined by the maximum stellar 
wind velocity.

\begin{figure}
\resizebox{\hsize}{!}{\includegraphics{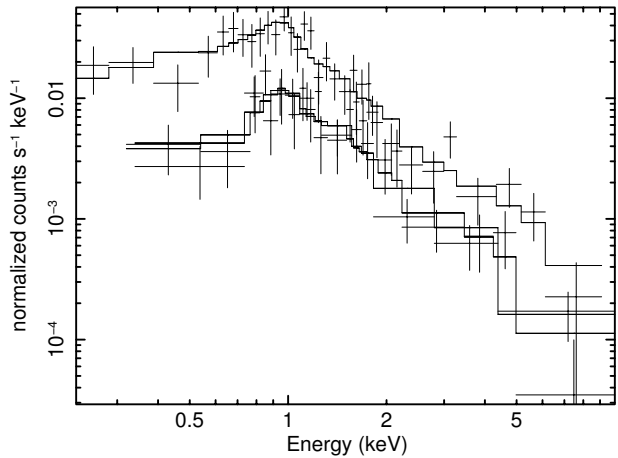}}
\caption{\XMM\ PN (upper curve), and MOS1 and MOS2 (lower curves) 
spectra of HR\,7355 with error bars corresponding to 3$\sigma$ with 
the best fit thermal plus power-law model (solid lines). The model 
parameters are shown in {Table~\ref{par_xray}}. 
}
\label{fig_xray}
\end{figure}

\begin{figure*}
\resizebox{\hsize}{!}{\includegraphics{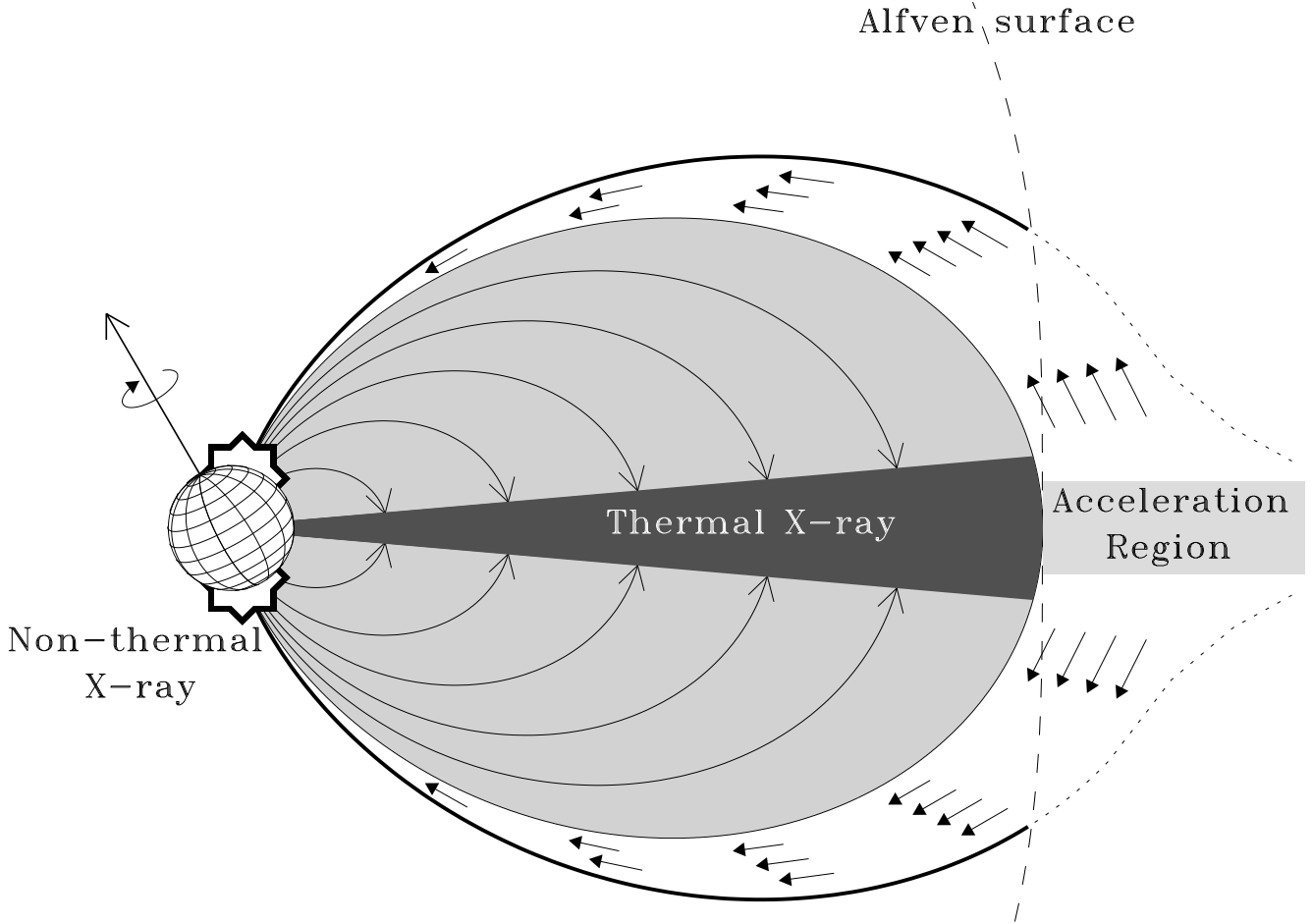}}
\caption{Meridian cross section of the model of 
the HR\,7355 magnetosphere,
that summarize where the radio and the X-ray emission originate.
The wind streams, arising from the two opposite stellar hemispheres,
collide  at the magnetic equator (dark grey area) 
giving rise to plasma heating and consequently X-ray emission.
The colder thermal plasma accumulates at the higher magnetic latitudes (grey 
area),
this plasma do not emit at X-ray but have a crucial role to explain the 
observed 
rotational modulation of the radio emission. Outside the Alfv\'{e}n surface 
the co-rotation breakdown take place
with a consequent formation of current sheet systems that are site of 
non-thermal 
electron injection.
The relativistic electrons that propagate toward the stellar surface 
(schematically represented by the small arrows)
radiate at the radio regime by the incoherent gyro-synchrotron emission 
mechanism.
The fraction of non-thermal electrons that impact with the stellar surface emit 
auroral X-ray emission, manifested in the X-ray spectrum by the non-thermal 
X-ray component.
}
\label{sezione_x}
\end{figure*}

In the  analysis presented in Sect.\,\ref{sec:radiodiagnostic}, 
we assumed two distinct wind velocities, 500\,km\,s$^{-1}$ and 
1000\,km\,s$^{-1}$, that encompass values plausible for a main 
sequence B-type star \citep{prinja89,oskinova_etal11,kritcka14}.
Using Eq.\,10 from \citet{ud-doula_etal14}, we estimate the
maximum plasma temperatures that can be produced via a magnetically confined
wind shock for these two wind speeds; these temperatures are
respectively: 3.5\,MK and 14\,MK -- significantly lower than that deduced
from the X-ray spectral analysis.

This led us to conclude that the assumption of the hard part 
of X-ray spectrum being produced by the hot thermal plasma 
is not realistic.  Therefore, as a next step, 
we attempted to fit the HR\,7355 spectra
with an absorbed power-law model, however no satisfactory fit could
be obtained. Finally, we fit the observed X-ray spectrum by 
combining thermal and power-law models.  
{The resulting fit, corresponding to the power-law plus thermal model,  is shown in Fig.~\ref{fig_xray}.}
A high-quality fit with reduced $\chi^2 = 0.725$
for 89 degrees of freedom was obtained. Based on spectral
fitting, the 2T thermal model has no preference over a model that
combines thermal and power-law (non-thermal) components.  The
model fit parameters are shown in Table~\ref{par_xray}.  The
temperature of the thermal X-ray plasma in this combined model, 
$\approx 10$ MK, is easier to reconcile with a typical wind 
velocity of a B2V star. A more complex model involving two 
temperatures plus a power-law, can also be fit to the observed spectra, 
yielding marginally better fitting statistics, however we choose 
the simplest models.

To investigate on the origin of the power-law X-ray component in
the HR\,7355 spectrum, it is useful to consider the Sun.  
The solar flare X-ray spectrum usually shows a hard X-ray component 
with a power-law energy distribution \citep{hudson_ryan95}, well explained 
as bremsstrahlung from a non-thermal electron population \citep{brown_71}.
For the non-thermal bremsstrahlung emission, the observed X-ray
spectral index ($\alpha$) is related to the spectral index ($\delta$)
of the injected non-thermal electrons.  For a thick-target
bremsstrahlung emission, $\delta$ and $\alpha$ are related as follow:
$\delta = \alpha +1$.
In the case of HR\,7355, if we assume that its power-law X-ray
emission is generated by the impact with the stellar surface of the
same non-thermal electron population responsible of the gyro-synchrotron
radio emission, it is possible to estimate a value $\delta=2.7$ of
the spectral index of these energetic electrons. 
The simulation of the radio emission of HR\,7355 indicates that the non-thermal
electrons have a spectral index of $\delta=2$--2.5.
The above range of values is not quite consistent with 
the value derived by the spectrum of the X-ray photons, but is however close.
This suggests that the thick-target bremsstrahlung 
emission from a non-thermal electron population that impacts
with the stellar surface is a plausible explanation for
the origin of the power-law component
detected in the X-ray spectrum of HR\,7355.
{Similarly to the Solar case, it is likely that 
not only electrons, but also protons are accelerated in HR\,7355. 
In some solar flares the observable effects of these two 
different populations of high energy particles have been 
recognized in the X- and the gamma-ray domain. 
The electrons radiate hard X-rays by thick-target bremsstrahlung emission,
whereas the protons, that interact with the ions at the stellar surface, 
radiate gamma-rays (see \citealp{aschwanden02} and references here reported). 
These effects, if present, will open a new 
high energy observational window to the hot magnetic stars.
}

The mechanism responsible for this power-law X-ray spectral component 
has an important difference with the case of solar flares.  During a solar
flare, the energetic electrons are impulsively injected, whereas
for hot star magnetospheres, the non-thermal electrons are
continuously accelerated.  The latter mechanism is similar to the
auroras from the magnetized planets in the solar system.  Thus, we 
suggest that the X-ray emission from HR\,7355 is physically analogous 
to the X-ray from Jupiter's aurora as measured by \XMM\ in November
2003. The Jovian X-ray spectrum has been modeled using a combination
of thermal and power-law components \citep{branduardi-raymont_etal_07}.
The power-law X-ray component dominates at the high energies of
the \XMM\ spectral range (X-ray photons with energies higher then 2
keV), and is explained in terms of bremsstrahlung 
emission of the precipitating
electrons with energies of $\approx 100$ keV
\citep{branduardi-raymont_etal_08}.  This high-energy electron
population is generated far from the planet (20--30 Jupiter radii),
in the co-rotation breakdown region of Jupiter's magnetosphere.

The mechanism we propose for the non-thermal electron acceleration 
in hot magnetic stars resembles the acceleration of high-energy electrons 
in Jupiter's atmosphere.  For the stellar magnetospheres, the co-rotation background
region coincides with the equatorial current sheet outside the
Alv\'{e}n radius (see Fig.~\ref{modello}).  The non-thermal electrons
are responsible for the gyro-synchrotron radio emission
of HR\,7355 and have a power-law energy distribution with a low-energy
cutoff at 100\,keV.  These particles can be, also,
responsible for the power-law component in the HR\,7355 X-ray
spectrum, as a consequence of the bremsstrahlung at the stellar
surface.  In this case, the bremsstrahlung X-ray emission arises
from an annular region around the pole.  The existence of a well
defined spatial location of the hard X-ray source region could
produce a smooth modulation of the X-ray emission as the star
rotates.  Unfortunately the short archival \XMM\ observations 
did not sample the stellar rotation period, so no conclusions 
can be drawn yet about the variability of the power-law component 
in the HR\,7355 X-ray spectrum. We intend to remedy this observational 
shortcoming in future.

In Figure~\ref{sezione_x} we present the model sketch that may 
explain simultaneously the radio and the 
X-ray emission of HR\,7355.
The thermal X-ray component arises from the hot plasma that is
shocked by the impact
between stellar wind flows from opposing stellar hemispheres.
The colder thermal plasma that accumulates in the the inner-magnetosphere
does not significantly contribute to the X-rays, but explains
the rotational modulation of the radio emission.
The stellar radio emission originates from a non-thermal electron population moving inside
the magnetic cavity that is defined by the field lines that
intercept the magnetic equator outside the Alv\'{e}n radius, coinciding 
with the magnetospheric regions where 
the co-rotation breakdown take place, and where the current sheet systems 
accelerate the electrons up to relativistic energies.
These precipitating non-thermal electrons could give rise to stellar auroral signatures,
like the non-thermal X-ray emission from the polar caps.

\section{Is there auroral radio emission from HR\,7355?}
\label{sec:aur}

The non-thermal electron population injected in the stellar magnetosphere
close to the Alfv\'{e}n radius, 
and which moves toward the star, is responsible for the incoherent 
gyro-synchrotron radio emission of HR\,7355. 
The electrons with velocities almost parallel with the magnetic field lines
(i.e., low pitch angle)  
can deeply penetrate within the stellar magnetosphere,
and are responsible for the X-ray auroral spectral features,
as discussed in Sec.~\ref{sec_xray}.
Non-thermal electrons that impact the stellar surface are lost to the magnetosphere.
As a consequence, 
the distribution of the non-thermal electrons reflected outside (by the magnetic mirroring) will be deprived
of electrons with quite low pitch-angle values. This is a suitable condition
to develop the unstable electron energy distribution, known as the loss-cone distribution
\citep{wu_lee79,melrose_dulk82}. 
This inverted electron velocity distribution 
can trigger the coherent Electron
Cyclotron Maser (ECM) emission mechanism.

\begin{figure}
\resizebox{\hsize}{!}{\includegraphics{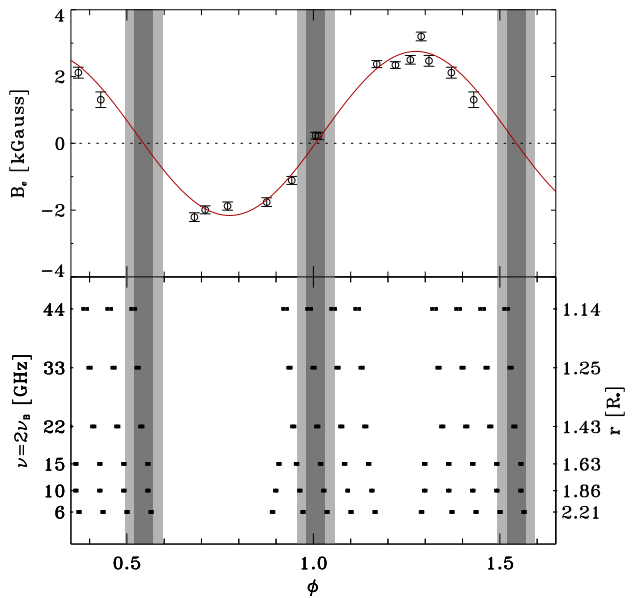}}
\caption{
Top panel: effective magnetic field curve of HR\,7355, are
from \citet{rivinius_etal10} and \citet{oksala_etal10}, a sinusoidal
fit of the data is superimposed.  Bottom panel: phase coverage of
the radio observations.
The y-axis on
the right indicates the distances of the layers where the observing
frequencies coincide with the second harmonic of the local
gyro-frequency.  Gray areas represent a phase interval that is 5\% of
the full rotation period; the light-gray areas represent
an interval of $\Delta \phi = 10$\% around the phases where the
effective magnetic field of HR\,7355 is null.}
\label{fase_coverage}
\end{figure}

The ECM amplifies the extraordinary magneto-ionic mode, producing
highly circularly polarized radiation ($\approx 100\%$), at frequencies 
close to the first few harmonics
of the local gyro-frequency ($\nu_{\mathrm B} = 2.8 \times 10^{-3}B/{\mathrm {Gauss}}$ GHz);
however, the fundamental harmonic is probably suppressed by gyro-magnetic 
absorption 
effects \citep{melrose_dulk82}.
The ECM is the process that generates the broad-band auroral radio emission
of the magnetized planets of the solar system \citep{zarka98}, 
of the two magnetic stars CU\,Vir \citep{trigilio_etal00, trigilio_etal08, trigilio_etal11, ravi_etal10, lo_etal12} 
and HD\,133880 \citep{chandra_etal15},
and, 
of the Ultra Cool Dwarfs
\citep{berger02, 
burgasser_putman05, 
antonova_etal08, 
hallinan_etal08, 
route_wolszczan12, 
route_wolszczan13, 
williams_etal15, 
kao_etal16}. 

The auroral radio emission arises from the thin density-depleted magnetic 
cavity 
related to the auroral oval at the polar caps.
This kind of coherent radio emission is 
efficiently amplified within a narrow beam pattern
tangentially directed along the cavity wall (i.e., the laminar source model, 
\citealp{louarn_lequeau96a,louarn_lequeau96b}),
giving rise to a radio light house effect.
In the case of the Earth Auroral Kilometric Radiation (AKR), this highly 
directional coherent radiation is upward refracted by the
dense thermal plasma trapped outside the auroral cavity \citep*{mutel_etal08,menietti_etal11}.

The auroral radio emission from CU\,Vir was first discovered at 1.4
GHz as intense ($\approx 1$ order of magnitude brighter then the
incoherent radio emission) radio pulses that were 100\% circularly
polarized \citep{trigilio_etal00}.  This amplified emission from
CU\,Vir has also been detected at 2.5 GHz \citep{trigilio_etal08},
and at 600 MHz \citep{stevens_george10}.  The pulse arrival times
are observed to be a function of the observing frequency
\citep{trigilio_etal11}, a signature of frequency-dependent refractive
effects suffered by the CU\,Vir auroral radio emission.

The radio pulses, related to the auroral radio emission from a 
dipole-shaped magnetosphere, 
occur when the magnetic dipole axis lies in the plane of the sky
(cross-over phases), with a duration of 5--10\% of the rotational
period.  The features of the auroral radio emission arising from
the CU\,Vir magnetosphere has been successfully modeled using a
simple dipole shape \citep{leto_etal16}.

Our source HR\,7355 has a stellar geometry, rotation, and
magnetic dipole obliquity suitable for detection of its auroral
radio emission.  In fact, as evident from the top panel
of Fig.~\ref{fase_coverage}, the curve for the magnetic field
shows a change of net
polarity twice per rotation.  The phase coverage of the 
radio observations presented in this paper has good sampling
at the cross-over phases at each observing frequency (see
bottom panel of Fig.~\ref{fase_coverage}).  At the frequencies
ranging from 6 to 44 GHz, the radio measurements of HR\,7355 do
not show any hint of auroral radio emission.


As previously explained, the stellar auroral radio emission at a
given frequency originates in the magnetospheric regions where the
second harmonic of the local gyro-frequency is close to the observing
frequency.  In the case of CU\,Vir ($B_{\mathrm p}=3800$ [Gauss],
\citealp{kochukhov_etal14}), emission from the ECM
arises from magnetospheric layers located between 
$\approx 1$ and $\approx 2.3$ stellar radii above the stellar surface.
Scaling to the polar strength of HR\,7355 ($B_{\mathrm p}=11600$
[Gauss], \citealp{rivinius_etal13}), the selected observing bands,
tuned at the frequencies of the second harmonic of the gyro-frequency,
{arise from magnetospheric layers with $r$ ranging between 1.14 and 2.21 R$_{\ast}$
(see the right $y$-axis of the bottom panel of Fig.~\ref{fase_coverage}),
that correspond to layers that are located 
at heights lower than $\approx 1.2$ stellar radii
from the surface of the star.}
On the other hand, from the model
simulations of the radio light curves of HR\,7355 (this paper) and
CU\,Vir \citep{leto_etal06}, we note that the thermal plasma trapped
within the magnetosphere of HR\,7355 has a higher density as compared
to CU\,Vir.  The auroral radio emission arising from the deeper
magnetospheric layers of HR\,7355 may suffer absorption effects.
Up to now, we have no radio measurements at frequencies
below 6 GHz, corresponding to layers further out in radius
where conditions would be more conducive for the detection of auroral
radio emission from HR\,7355.  Consequently, it is not possible at
this time to draw firm conclusions regarding the production of
auroral radio emission at HR\,7355.

\section{Summary and conclusions}
\label{sec:sum}

In this paper we have presented an
extensive analysis of the rigidly rotating magnetosphere of the 
fast rotating B2V star, HR\,7355.
This study has been made using radio (VLA) and X-ray (\XMM) observations.

The radio measurements of HR\,7355 cover a large frequency range, from 6 to 44 GHz.
The total (Stokes I) and the circularly polarized (Stokes V) flux density are variable.
The radio data have been phase folded using the stellar rotation period,
demonstrating that the radio variability is a consequence of the rotation.
The rotational phases have been well sampled, allowing us
to build multi-wavelength radio light curves separately in Stokes I and V.
Modeling of the stellar radio emission, 
using a 3D model that simulates the radio emission from a rigidly 
rotating stellar magnetosphere shaped by a simple dipole
\citep{trigilio_etal04,leto_etal06},
allows us to constrain the physical conditions in the magnetosphere of HR\,7355.
As result of the present analysis, the wind mass-loss rate of HR\,7355 has been indirectly derived. 
Independently, we obtained constrains on 
mass-loss rate from the analysis of archival UV spectra of HR\,7355 by means of 
the non-LTE stellar atmosphere model PoWR. The radio and UV values of mass-loss 
rate are in good agreement, and are in accord with estimates of mass-loss 
rates derived from the UV spectra of other stars with similar spectral types.

The average radio luminosity of HR\,7355 is about 
$10^{18}$\,erg\,s$^{-1}$\,Hz$^{-1}$, in the range 6--44 GHz,
making it one of the brightest radio sources among the class of the MCP stars
(mean radio luminosity $\approx 10^{16.8}$ [ergs s$^{-1}$ Hz$^{-1}$]
\citealp{drake_etal87,linsky_etal92}).
To investigate further, the magnetosphere of HR\,7355 was compared 
with that of CU\,Vir, another magnetic star studied with the same 
modeling approach.
The comparison reveals that both these stars are characterized by 
centrifugal magnetospheres but, the same time, have significant differences.
The CM of HR\,7355, as normalized to the stellar radius is larger,
and the regions where the non-thermal electrons are generated
are characterized by a stronger local magnetic field strength, 
with respect to the case of CU\,Vir,
with a consequent effect on the radio brightness of the two stars.

The analysis presented in this paper allows us to estimate the 
average physical conditions of the thermal plasma confined within the 
magnetospheric region. 
Absorption effects by the trapped plasma influences the
emerging stellar radio emission and plays a key role for the modeling
of the radio light curves. 
The measured thermal X-ray emission from HR\,7355 could be explained as a 
consequence of the shock heating of the colliding wind streams arising from the 
two opposite stellar hemispheres.

The fit to the X-ray spectrum of HR\,7355 suggests a presence of
a non-thermal X-ray component described by a power law. The spectral index of 
the non-thermal X-ray
photons is compatible with the thick target bremsstrahlung emission generated by
the non-thermal electron population, which are responsible for the observed 
radio emission, that impact with the stellar surface close to the polar caps.
This could be the signature for auroral X-ray emission from HR\,7355.

Stellar rotation can lead to greater X-ray emissions than predicted by the
scaling laws in the framework of the MCWS model \citep{bard_townsend15}.
However, \cite{ud-doula_naze15} point out that even taking rotation into 
account, there are some hot magnetic stars that are too bright in the X-ray 
band, one of them being HR\,7355. 
Our new model suggests that auroral 
X-ray emission is a likely additional mechanism that increases X-ray production 
and can account for the strong X-rays from HR\,7355. We speculate that auroral 
mechanism operates in other hot magnetic stars that display hard and bright 
X-ray emission.

The auroral phenomenon also gives rise to features at radio wavelengths,
such as coherent pulses with $\approx$ 100\% circular polarization
that occur at predictable rotational phases.
This radio lighthouse phenomena has been recognized in the
magnetized planets of the solar system,
among some ultra cool dwarfs, and in two
hot magnetic stars, with the prototype CU\,Vir. 
For the stellar auroral radio emission to be detectable,
the stellar magnetic dipole must to be oriented with the axis
lying in the plane of the sky (null effective magnetic field).
The stellar geometry of HR\,7355
is favorable for detecting this coherent emission, 
in fact the magnetic field curve inverts the sign twice per period.
We do not however find any signature of auroral radio emission from HR\,7355, at least in the frequency range 6--44 GHz.  
For this frequency range, we suggest that
the auroral radio emission originates deep
inside the stellar magnetosphere and is strongly absorbed.
New observations at lower frequencies, corresponding to less deep layers, 
could reveal whether auroral radio emissions are in
fact produced the magnetosphere of HR\,7355.

We want to underscore that
the synergistic radio and X-ray analysis is
a powerful combination that can led to strong constraints for the
stellar magnetospheric conditions of hot magnetic stars.
From the results of the radio modeling simulations and the
X-ray spectral analysis of HR\,7355,
we have been able to outline a physical
scenario that simultaneously explains features detected at
opposite ends of the source spectrum.

\section*{Acknowledgments}
We thank the referee for their very useful remarks that helped to improve the paper. 
The National Radio Astronomy Observatory is a facility of the National Science 
Foundation operated under cooperative agreement by Associated Universities, Inc.
This work has extensively used  the NASA's Astrophysics Data System, and the 
SIMBAD database, operated at CDS, Strasbourg, France. This publication used  
data products provided by the XMM-Newton Science Archive. LMO acknowledges 
support by the DLR grant 50\,OR\,1302.

\end{document}